# Dynamics of a dielectric droplet suspended in a magnetic fluid in electric and magnetic fields

Arthur Zakinyan, Elena Tkacheva, Yury Dikansky

Department of Physics, Stavropol State University, 1 Pushkin St., 355009 Stavropol, Russian Federation

Address correspondence to A. Zakinyan:
postal address is presented above,
e-mail: zakinyan.a.r@mail.ru

**Abstract.** The behavior of a microdrop of dielectric liquid suspended in a magnetic fluid and exposed to the action of electric and magnetic fields is studied experimentally. With increasing electric field, the deformation of droplets into oblate ellipsoid, toroid and curved toroid was observed. At the further increase in the electric field, the bursting of droplets was also revealed. The electrorotation of deformed droplets was observed and investigated. The influence of an additional magnetic field on the droplet dynamics was studied. The main features of the droplet dynamics were interpreted and theoretically examined.

## 1. Introduction

Electrohydrodynamic deformation and burst of liquid drops suspended in another immiscible liquid are well described and investigated experimentally [1–5], theoretically [3–8], and computationally [9–11]. Briefly summarizing the results of previous works, it can be concluded that depending on certain system parameters the drop can take the equilibrium shape of oblate or prolate ellipsoid and can burst into smaller drops under the action of electric field. The electric field can also bring into rotation the liquid drops as it was reported in [12, 13]. The droplets electrorotation studied in [12, 13] is quite similar in nature to the electrorotation of solid particles suspended in a liquid (see [14] for a review). While the dynamics and deformation of freely suspended droplets has been well analyzed and observed, the deformation of droplets confined in the thin film geometry still remains relatively unstudied. In addition, some new peculiarities of the droplet behavior can appear in a system where the droplet relaxation time is large. The investigation of the mentioned problems is the subject of the present work.

Magnetic fluid is an artificially synthesized fluid with noticeable magnetic properties. It represents a colloidal suspension of ultra-fine ferro- or ferri-magnetic nanoparticles suspended in a carrier fluid. The action of a magnetic field on magnetic fluids was studied intensively, both theoretically and experimentally, in many works. In particular, the action of a uniform magnetic field on magnetic fluid drops has been much studied. The review of these works is presented in [15]. The obtained results demonstrate a wide analogy between the behavior of a liquid drop in an electric field and the behavior of a magnetic fluid drop in a magnetic field. But, unlike an electric field, a uniform magnetic field always leads to the stretching of both a magnetic fluid drop in a nonmagnetic liquid and a nonmagnetic drop in a magnetic fluid.

Because the droplets of magnetic fluid and the nonmagnetic droplets suspended in a magnetic fluid can be deformed in both magnetic and electric fields, the simultaneous effect of electric and magnetic fields on a droplet may lead to behaviors much more complicated and varied. Some peculiarities of these behaviors have been studied experimentally in [16] and theoretically in [17]. In the present work we report on the further development of such studies, here we investigate the dynamics of a single droplet of a dielectric liquid suspended in a magnetic fluid under the action of electric and magnetic fields.

## 2. Experiments

In our experiments we used a kerosene-based magnetic fluid with dispersed magnetite nanoparticles of about 10 nm diameter stabilized with oleic acid. The properties of the magnetic fluid are: dynamic viscosity is $\eta_e = 30$ mPa·s, conductivity is $\sigma_e = 1.3 \cdot 10^{-6}$ S/m, dielectric constant is $\varepsilon_e = 5.2$, magnetite volume fraction is 13% and saturation magnetization is 55.4 kA/m. Liquid caoutchouc immiscible with the magnetic fluid was selected as the dielectric liquid. Its density is $\rho = 725$ kg·m$^{-3}$, dynamic viscosity is $\eta_i = 1.5$ Pa·s, conductivity is $\sigma_i = 10^{-12}$ S/m and dielectric constant is $\varepsilon_i = 2.3$. The reasons to use liquid caoutchouc are that its viscosity is comparatively high and the interfacial tension at the interface between it and the magnetic fluid is very low ($\gamma = 6.7 \cdot 10^{-6}$ N/m). The interfacial tension was determined by the retraction of the deformed drop method proposed in [18]. The experimental sample was prepared by mechanical mixing of a small volume of liquid caoutchouc with the magnetic fluid. The radii, $R$, of droplets of the emulsion obtained in this way are varying from 5 to 30 μm. Since the size of magnetic nanoparticles of magnetic fluid is much smaller than the sizes of nonmagnetic liquid caoutchouc droplets suspended in the magnetic fluid, the fluid can be considered as a continuous liquid magnetizable medium. No stabilizing agents were used in the preparation of the emulsion. The volume fraction of the suspended droplets was about 0.01. Owing to the small volume fraction we could explore the behavior of a single droplet and neglect the effects of droplets interaction.

Droplet behavior was experimentally studied by observations with an optical microscope, which was placed between Helmholtz coils generating a constant uniform magnetic field in the space where the cell containing a sample was located. Fig. 1 shows schematically the experimental setup, which was used for the droplet behavior investigation. We used two different types of cells. The first cell (presented at right-bottom in Fig. 1) was assembled of two rectangular flat glass plates covered with a transparent conducting stannic oxide coating. A fluoroplastic film 20–30 μm thick with a circular hole in the center was placed between the plates to set the distance between electrodes. The hole was filled with the studied emulsion sample. A variable voltage applied to the plates generated the ac electric field between them. The most droplets of the sample were approximately of the same size as the distance between electrodes, some droplets were smaller and some large droplets were slightly compressed by the plates of electrodes. For detailed observation of the droplet dynamics peculiarities the second cell representing a microscope slide with two rectangular metal plates glued on the slide surface was also used (presented at left-bottom in Fig. 1). The distance between edges of metal plates was 1 mm and their thickness was 0.1 mm. The space between edges of metal plates was filled with an emulsion, and a voltage applied to the plates generated an electric field between them. The first cell served for viewing the events through a microscope along the electric field direction and for studying the behavior of a droplet confined in the thin film geometry. The second cell served for viewing perpendicularly to the field direction and for studying the behavior of a freely suspended droplet. It should be noted that the main difficulty encountered in the study of the behavior of bodies in a magnetic fluid is that the magnetic fluid is an opaque medium, which complicates observations. However, relatively thin layers of a magnetic fluid (< 100 μm) are transparent enough to study the behavior of drops in such fluids. At first we studied the droplet behavior under the action of ac electric field and then under the simultaneous action of ac electric and dc magnetic fields. Because of low interfacial tension, under the action of comparatively weak electric and magnetic fields significant deformation of droplets can take place.

Using the first cell it was experimentally observed that at relatively low frequencies of electric field, the droplet is flattened taking the shape of oblate ellipsoid of revolution; at higher frequencies, the droplet stretches along the force lines of the electric field and becomes prolate ellipsoid. The deformation parameter, $D$, has been measured as a function of a frequency, $f$, and of a root-mean-square value, $\overline{E}$, of an alternating electric field ($D=(b-a)/(b+a)$, here $b$ is the semiaxis parallel to the field and $a$ is the semiaxis perpendicular to it). The obtained experimental dependencies of the deformation parameter on the electric field strength and frequency are shown in Figs. 2 and 3. It was found that $D$ linearly depends on $\overline{E}^2$ and rises with $f$ increasing. At some critical value of frequency, $D$ reverses a sign. It was observed that the action of a magnetic field leads to the stretching of droplets, therefore when a droplet is flattened in a low-frequency electric field, the deformation can be compensated by imposing an additional magnetic field directed in parallel with the electric field.

In a low-frequency range of electric field the oblate ellipsoids arise at comparatively low $\overline{E}$ values (Fig. 4b). With increasing electric field strength the shape of a confined droplet changes: a hole appears in the middle of the oblate droplet and it takes a toroidal shape (Fig. 4c). Upon a further rise in the electric field strength the droplet shape becomes distorted and forms a curved toroid (Fig. 4d). At the final stage, the droplet bursts into several smaller droplets as the electric field increases (Fig. 4e). These small droplets then begin to rotate. The evolution of the confined droplet shape with increasing electric field strength at a constant field frequency is shown in Fig. 4. The values of $\overline{E}$ at which the mentioned droplet configurations can appear depend on the field frequency. The experimentally obtained phase diagram showing the ranges of $\overline{E}$ and $f$ in which the different droplet configurations appear, is presented in Fig. 5a. The additional action of a dc magnetic field aligned with the electric field changes the droplet shape evolution. Particularly sufficiently strong magnetic field can lead to the disappearance of the toroidal configuration of a droplet and the phase of droplet shape distortion comes right after the oblate ellipsoid configuration. The phase diagrams illustrating the influence of an additional magnetic field on the droplet shape dynamics at two different values of a magnetic field strength ($H = 550$ and 920 A/m) are presented in Figs. 5b and 5c.

As it was already mentioned above, the electrorotation of solid particles and liquid drops suspended in a leaky dielectric liquid has been widely studied. The undertaken explorations show that the rotation of droplets under the action of electric field in our experiments is a phenomenon quite different from that observed in the previous works. In the above experiments with a confined droplet, the rotation can take place only for small droplets arisen from the disintegration of the initial drop. For better study of the electrorotation phenomenon the observations of the behavior of freely suspended droplets were carried out by means of the second cell. The observations show that in our case the electrorotation passes in the following way. It can be observed only in a low-frequency electric field when droplet flattening takes place. At first the droplet flattens as the electric field increases. When the droplet deformation reaches some critical value it begins to turn over and tends to be orientated by a major semiaxis to the electric field direction (droplet turns through 90°). After orientation the droplet restores to its original form of an oblate ellipsoid with a major semiaxis perpendicular to an electric field.

And this events sequence occurs over and over again. The rotation direction may be either clockwise or counterclockwise. The consecutive snapshots of the droplet showing the described process are presented in Fig. 6. It should be noted that when a droplet is suspended freely not only small droplets can rotate; the electrorotation of droplets of any size was observed in this case. Comparatively large droplets flattened in a low-frequency electric field and confined in thin film geometry cannot rotate because they are restricted by electrodes plates. The toroidal configuration was not observed for freely suspended droplets. In conclusion it may be said that the oblate deformation of a droplet is a necessary condition for the droplet rotation. On the contrary, the rotation of particles and droplets observed in the previous works was not conditioned by the deformation.

It can be supposed that the observed unusual electrorotation is caused by a specific properties of a studied system such as quite low surface tension and high viscosity of liquid of a droplet (shape relaxation time is large, ~ 5 s). The dependences of the droplet rotation frequency, $v$, on strength and frequency of the electric field have been measured (Figs. 7 and 8). For the purposes of detection of measurement errors three series of measurements were made. The results of all three series of measurements are presented in the figures. As is seen, the droplet rotation frequency grows with the electric field strength increasing. In a low-frequency range of an electric field the droplet rotation frequency grows too, but this dependence rapidly comes to saturation with the electric field frequency increasing. It was found that under the additional action of a constant magnetic field directed in parallel to the electric field the rotation frequency of a droplet grows. The obtained experimental dependence of the droplet rotation frequency on the magnetic field strength is presented in Fig. 9. But in the strong magnetic field the droplet rotation comes to a stop when the compensation of a droplet deformation by a magnetic field takes place.

When a droplet confined in a thin film bursts into an ensemble of rotating smaller droplets, these droplets repel each other in an electric field and tend to draw up a hexagonal structure. Diffraction light scattering by this structure has been studied. For this purpose a laser beam was transmitted through the emulsion layer (Fig. 10). The ratio of the radius of the first diffraction ring to the distance to diffraction image, $r/d$, has been measured as a function of an electric field strength and frequency; the results are shown in Fig. 10. As is seen, the radius of the diffraction ring grows with the electric field strength increasing, and decreases with increasing frequency. These results can be explained in the following way. The diffraction ring radius is inversely proportional to the structure period. The action of sufficiently strong electric field leads to the break-up of emulsion droplets, as a result the distance between the adjacent droplets diminishes with electric field increasing and the ring radius grows. With reference to Fig. 5, it can be seen that the electric field strength at which the droplet break-up takes place grows with the electric field frequency increasing. It can lead to the diminishing of the diffraction ring radius with increasing electric field frequency when the field strength is fixed. In summary it may be said that the diffraction experiment corroborates the results obtained by direct observations of the droplet dynamics.

## 3. Analysis and discussion

The measured droplet deformation has been compared with theoretical results of [5]. According to [5] the deformation parameter can be calculated by following expression

$$D = \frac{9\varepsilon_0 \varepsilon_e}{16\gamma} \Phi \overline{E}^2 R, \qquad (1)$$

$$\Phi = 1 - \frac{B(11\lambda + 14) + B^2[15(\lambda + 1) + q(19\lambda + 16)] + 15A^2\omega^2(1+\lambda)(1+2q)}{5(1+\lambda)[(2B+1)^2 + A^2\omega^2(q+2)^2]}$$

where $\varepsilon_0$ is the permittivity of free space, $A = \varepsilon_0 \varepsilon_e / \sigma_i$, $\omega = 2\pi f$, $q = \varepsilon_i / \varepsilon_e$, $\lambda = \eta_i / \eta_e$, $B = \sigma_e / \sigma_i$. The results of calculations are presented in Figs. 2 and 3. The comparison shows that there is a qualitative agreement between theory and experiment, but one can see that the results of measurements differ quantitatively from the calculations. The difference can be an effect of simplifying assumptions made in a first approximation theory [5]. Among these assumptions are the consideration of small deformations only, neglect of the surface conductance and the convection of charge, neglect of the effects due to diffuse ionic layers at the interface, neglect of the electrocapillarity effects, etc.

The appearance of toroidal shape of a droplet (Fig. 4*c*) is a quite complicated phenomenon for detailed analysis. As a hypothesis, it may be supposed that the appearance of a hole and the toroidal configuration of a droplet result from the flow development inside and outside the droplet. The flow pressure is maximal at the pole of an oblate droplet [5], at the same time the surface pressure is minimal at the pole, as a result the hole appears at the droplet pole under the action of the flow pressure. It seems that the possibility of rotation of a freely suspended droplet changes the flow pattern considerably, and the toroidal shape of a droplet does not appear in this case. The described distortion of a droplet shape in an electric field (Fig. 4*d*) can take place in thin film geometry only and was not observed for freely suspended droplet. It suggests that this effect is similar in nature to the well-known instability of magnetic fluid drops in a thin layer under the action of a perpendicular magnetic field (see [15] for a review).

We shall restrict our consideration to the effect of droplet electrorotation. The specific electrorotation described above can occur in a case when a droplet relaxation time is greater than a characteristic time of orientation of an ellipsoidal droplet along the electric field. Therefore, we will find and compare these times. The electric field produces torque acting on a droplet. According to [14], the electric torque, $M_e$, can be written as

$$M_e = \tfrac{1}{3}\pi a^2 b \varepsilon_0 \varepsilon_e (n_y - n_x) E_0^2 \operatorname{Re}(K_x K_y) \sin 2\alpha = M_e^0 \sin 2\alpha, \qquad (2)$$

$$K_{x,y} = \tfrac{1}{3}(\varepsilon_i^* - \varepsilon_e^*)/[\varepsilon_e^* + (\varepsilon_i^* - \varepsilon_e^*) n_{x,y}],$$

where $\varepsilon_{i,e}^* = \varepsilon_{i,e} + \sigma_{i,e}/j\omega\varepsilon_0$ is the complex dielectric constant, $j$ the imaginary unit, $E_0$ the electric field amplitude, $\alpha$ is the angle between the field direction and the ellipsoid major semiaxis, $n_{x,y}$ the depolarization factor given by

$$n_x = \frac{1+e^2}{e^3}(e - \operatorname{atan} e),\ n_y = (1 - n_x)/2,\ e = \sqrt{\frac{a^2}{b^2} - 1}\,.$$

We suppose that the droplet shape and its semiaxes are determined by (1) and are changeless during the orientation process. The viscous torque acts in opposition to the electric torque. According to [19], the viscous torque, $M_\eta$, can be written in the form

$$M_\eta = C\Omega, \qquad (3)$$

$$C = \eta_e V \frac{(n_y - n_x)\tfrac{N}{2}\left(\tfrac{a}{b} - \tfrac{b}{a}\right) - n'(a^2 + b^2)\left[1 + \tfrac{N}{2}\left(\tfrac{a}{b} + \tfrac{b}{a}\right)\right]}{(a^2 n_x + b^2 n_y) n'},$$

$$N = \frac{4abn'}{\left(\tfrac{a}{b} - \tfrac{b}{a}\right)^2 \left(1 - \tfrac{\eta_i}{\eta_e}\right)(a^2 n_x + b^2 n_y) n' - \left(\tfrac{a}{b} - \tfrac{b}{a}\right)\left(\tfrac{a}{b} n_x - \tfrac{b}{a} n_y\right) - 2abn'\left(\tfrac{a}{b} + \tfrac{b}{a}\right)},$$

$$n' = \frac{n_y - n_x}{a^2 - b^2}$$

where $V = \tfrac{4}{3}\pi a^2 b$ is a droplet volume, $\Omega = d\alpha/dt$.

The equation of rotary motion of a droplet has a form

$$I\frac{d^2\alpha}{dt^2} = M_e + M_\eta \qquad (4)$$

where $I = \tfrac{m}{5}(a^2 + b^2)$ is the moment of inertia of an oblate ellipsoid, $m = \rho V$ the droplet mass. Substitution of Eqs. (2) and (3) into Eq. (4) gives the differential equation of a droplet orientation in the form

$$d^2\alpha/dt^2 + 2\beta\, d\alpha/dt + \omega_0^2 \sin 2\alpha = 0 \qquad (5)$$

where $2\beta = -C/I$, $\omega_0^2 = -M_e^0/I$. The performed calculations show that the motion of studied droplets is inertialess and the inertial term (second derivative of an angle) of the equation of motion (5) can be neglected. In the approximation of inertialess motion both torques are equal and oppositely directed. On this basis we can obtain the equation of a droplet orientation in the form

$$t = \frac{C}{4M_e^0} \ln \frac{(1+\cos 2\alpha)(1-\cos 2\alpha_0)}{(1-\cos 2\alpha)(1+\cos 2\alpha_0)} \qquad (6)$$

where $\alpha_0$ is the initial value of $\alpha$. The results of calculation of the dependence (6) are presented in Fig. 11.

According to the small deformation theory [20], the evolution of $D(t)$ during the drop relaxation can be written as

$$D = D_0 \exp(-st/\tau), \qquad (7)$$

$$s = \frac{40(\eta_i/\eta_e + 1)}{(2\eta_i/\eta_e + 3)(19\eta_i/\eta_e + 16)}, \quad \tau = \eta_e R/\gamma$$

where $D_0$ is the initial droplet deformation defined by Eq. (1). The dependence $D(t)$ calculated by Eq. (7) is also presented in Fig. 11. The dependences $\alpha(t)$ and $D(t)$ were calculated at three different values of an electric field strength ($E_0$ = 170, 210 and 250 kV/m). As is seen, at low electric field, the orientation time is greater than a relaxation time. But with increasing electric field, orientation time becomes less than relaxation time. From this it follows that the rotation of a droplet can take place when electric field strength reaches some critical value in a complete concordance with that observed in experiments.

Evidently the increase in an electric field strength leads to the increasing of turning moment and the frequency of droplet rotation grows. The value of $\alpha = \pi/4$ corresponds to the instable equilibrium of a droplet, and $\alpha = 0$ is an asymptote of the function $\alpha(t)$. To estimate the time, $t$, of the droplet orientation we will assume that $\alpha_0 = \pi/4 - 10^{-5}$ rad and $\alpha = 10^{-5}$ rad, and substitute these values into Eq. (6). This gives us an opportunity to estimate the droplet rotation frequency by an approximate expression $v \sim 1/(2t)$. Fig. 7 shows the calculated dependence of the droplet rotation frequency on the electric field strength. As is seen, experimental results compare well with theoretical calculations. The electric torque acting on a droplet with fluid less conductive than that of the ambient fluid depends very weakly on the electric field frequency, and so the droplet electrorotation also should not depend on it. The observed dependence of a droplet rotation on an electric field frequency in a low frequency range can be caused by the screening of real electric field inside the cell by the electric double layers which disappear with electric field frequency increasing.

Additional action of a parallel magnetic field leads to an increase in a droplet rotation frequency. The magnetic field produces torque acting on a droplet. The magnetic torque, $M_m$, acting on a nonmagnetic oblate ellipsoid immersed in a magnetic medium can be written as [21]

$$M_m = \frac{\mu_0 (\mu-1)^2 (n_y - n_x) V H^2}{2(\mu + (1-\mu) n_x)(\mu + (1-\mu) n_y)} \sin 2\alpha \qquad (8)$$

where $\mu_0 = 4\pi \cdot 10^{-7}$ H/m, $\mu$ is the magnetic permeability of magnetic fluid. The magnetic torque (8) is parallel to the electric torque (2). As a consequence under the additional action of magnetic field a total turning moment increases, and the droplet rotates faster. However the magnetic field deforms the droplet and its shape cannot be determined by Eq. (1) in this case. It leads to the difficulties in the analysis of the droplet rotation in electric and magnetic fields, and we will not discuss it here at greater length.

## 4. Conclusions

Thus, the presented study demonstrates some new peculiarities of the droplet behavior such as the appearance of toroidal shape and specific rotation under the action of external fields. The freely suspended droplet and droplet confined in thin film geometry under the simultaneous action of electric and magnetic fields have been studied.

It should be noted that the study and design of new composite material systems based on magnetic fluids currently attracts much attention owing to wide potential applications of such systems. One of new materials based on magnetic fluid is a magnetic fluid emulsion, disperse system composed of two liquid phases one of which is a magnetic fluid. In most studied cases the droplets of such emulsions hold the spherical shape and align in chain-like aggregates under the action of magnetic field, see e.g. [22, 23]. Much less attention has been paid to the study of magnetic fluid emulsion with deformable droplets. The deformation effect on the emulsion properties has been considered only in [24, 25]. In this study we show that the structure of synthesized magnetic fluid emulsions can considerably depend on the magnetic and electric fields. It is obvious that the structural organization in such systems can lead to the appearance of specific features in their macroscopic properties. The results of the present study can also be used in the design of devices in which magnetic fluid emulsions are used.

**Acknowledgments**

This work was supported by Russian Foundation for Basic Research (project No. 10-02-90019-Bel_a) and also by Ministry of education and science of the Russian Federation in scientific program "Development of Scientific Potential of Higher School".

**Figure Captions**

Fig. 1. Sketch of the experimental setup: *1* – sample cell; *2* – Helmholtz coils; *3* – optical microscope; *4* – digital video camera; *5* – microscope slide; *6* – metal plates; *7* – cover glass; *8* – transparent glass electrodes; *9* – fluoroplastic film. In the experiments electric and magnetic fields were parallel.

Fig. 2. Droplet deformation parameter vs. the electric field strength at three different values of field frequency: *1* – $f = 20$ Hz; *2* – $f = 1$ kHz; *3* – $f = 5$ kHz ($R = 28$ μm). Dots are experiments; lines are calculations.

Fig. 3. Droplet deformation parameter vs. the electric field frequency at three different values of field strength: *1* – $\bar{E} = 160$ kV/m; *2* – $\bar{E} = 175$ kV/m; *3* – $\bar{E} = 190$ kV/m ($R = 28$ μm). Dots are experiments; lines are calculations.

Fig. 4. Confined dielectric droplet at different values of electric field: $a$ – $\overline{E} = 0$; $b$ – $\overline{E} = 150$ kV/m; $c$ – $\overline{E} = 200$ kV/m; $d$ – $\overline{E} = 315$ kV/m; $e$ – $\overline{E} = 420$ kV/m ($f = 8$ Hz). Electric field is directed perpendicularly to the plane of a figure.

Fig. 5. Experimental phase diagrams showing the ranges of $\overline{E}$ and $f$ in which the different droplet configurations appear: ($a$) – $H = 0$; ($b$) – $H = 550$ A/m; ($c$) – $H = 920$ A/m. The numbers denote configurations: $1$ – oblate ellipsoid; $2$ – toroid; $3$ – curved toroid; $4$ – droplet burst and electrorotation; $5$ – intense electrohydrodynamic flows in the entire sample. Lines are the approximation of experimental data.

Fig. 6. The consecutive snapshots showing the repetitive process of electrorotation of a freely suspended droplet.

Fig. 7. Droplet rotation frequency vs. the electric field strength ($R = 5$ μm, $f = 8$ Hz). Dots are experiments; line is calculations.

Fig. 8. Droplet rotation frequency vs. the electric field frequency ($R = 6$ μm, $\overline{E} = 520$ kV/m).

Fig. 9. Droplet rotation frequency vs. the magnetic field strength ($R = 4$ μm, $f = 8$ Hz, $\overline{E} = 400$ kV/m).

Fig. 10. Measured ratio of the radius of the first diffraction ring to the distance to diffraction image vs. the electric field strength (open symbols, $f = 8$ Hz) and frequency (filled symbols, $\overline{E} = 300$ kV/m). Lines are the approximation of experimental data.

Fig. 11. Calculated dependences $\alpha(t)$ (solid lines) and $D(t)$ (dashed lines) at different values of electric field: $1 - E_0 = 170$ kV/m, $2 - E_0 = 210$ kV/m, $3 - E_0 = 250$ kV/m ($f = 20$ Hz, $R = 25$ μm).

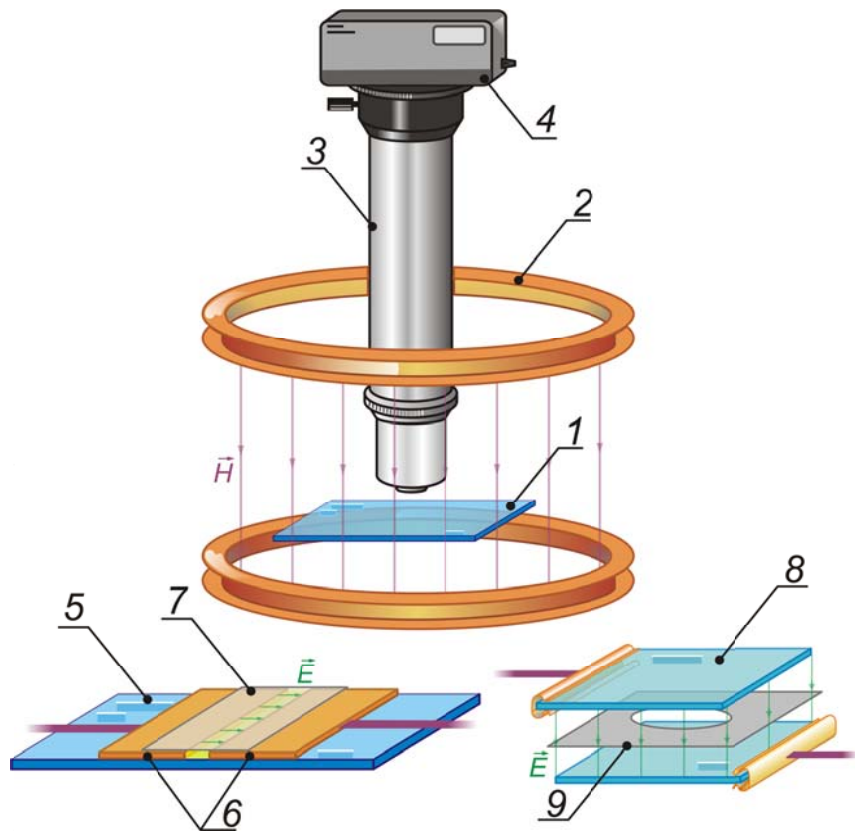

FIG. 1.

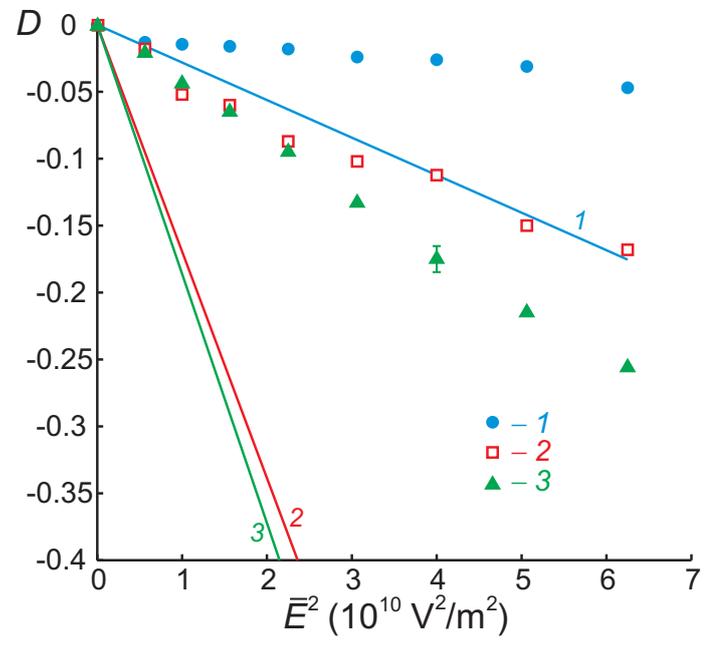

FIG. 2.

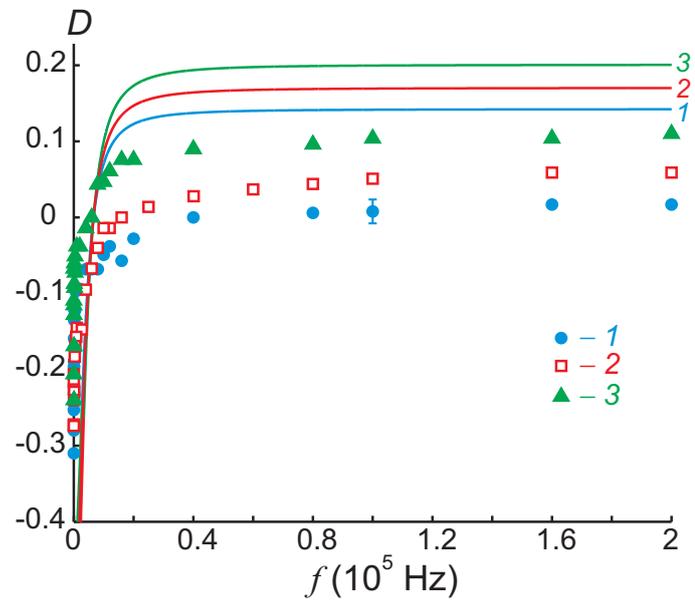

FIG. 3.

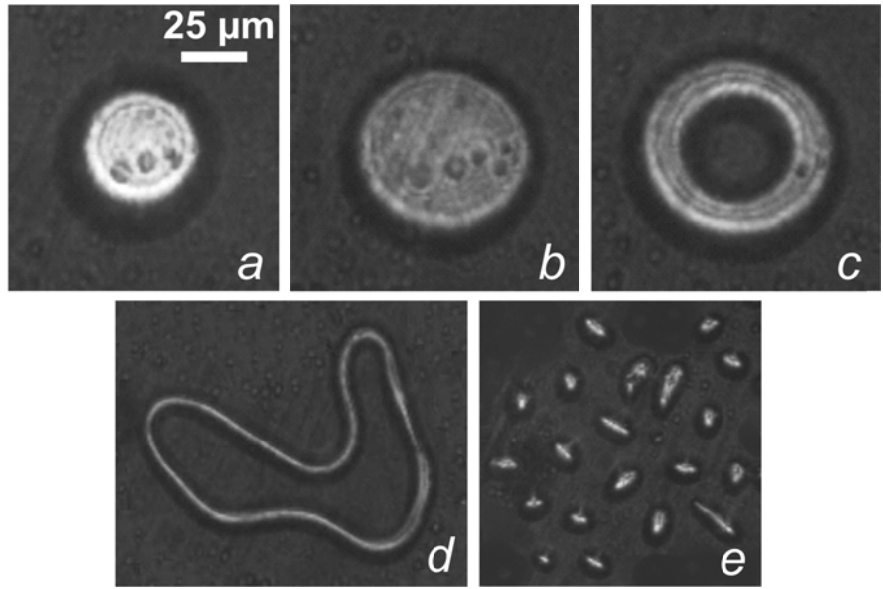

FIG. 4.

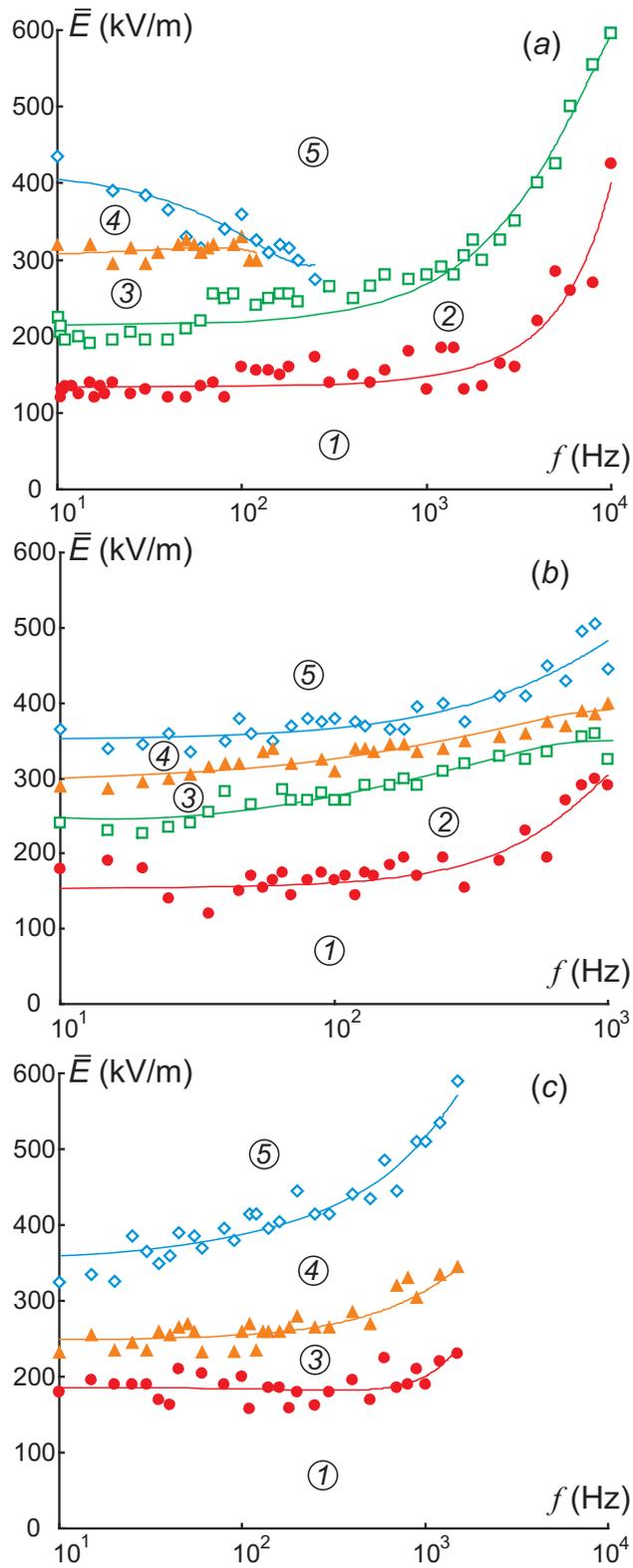

FIG. 5.

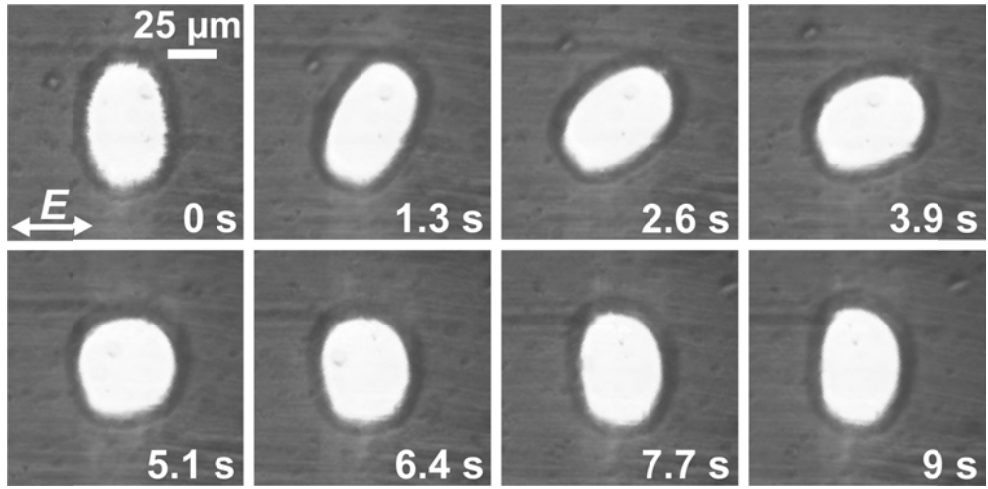

FIG. 6.

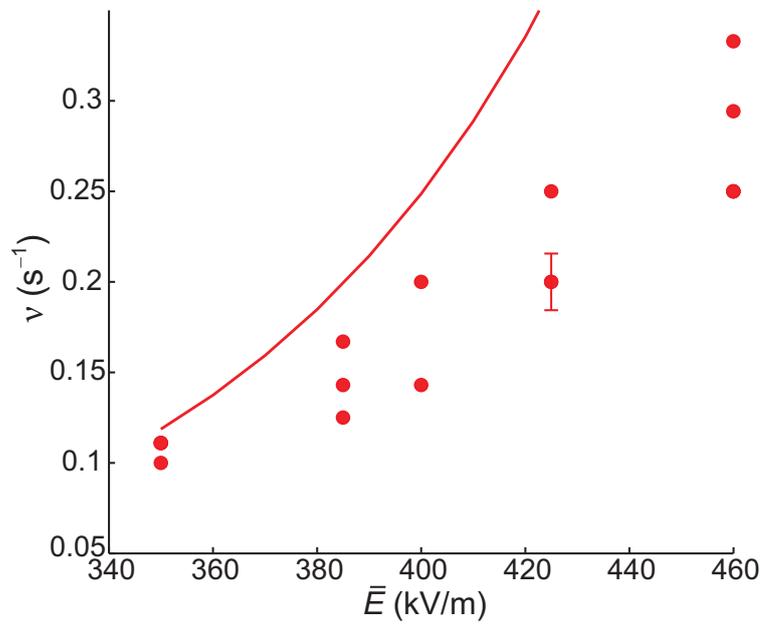

FIG. 7.

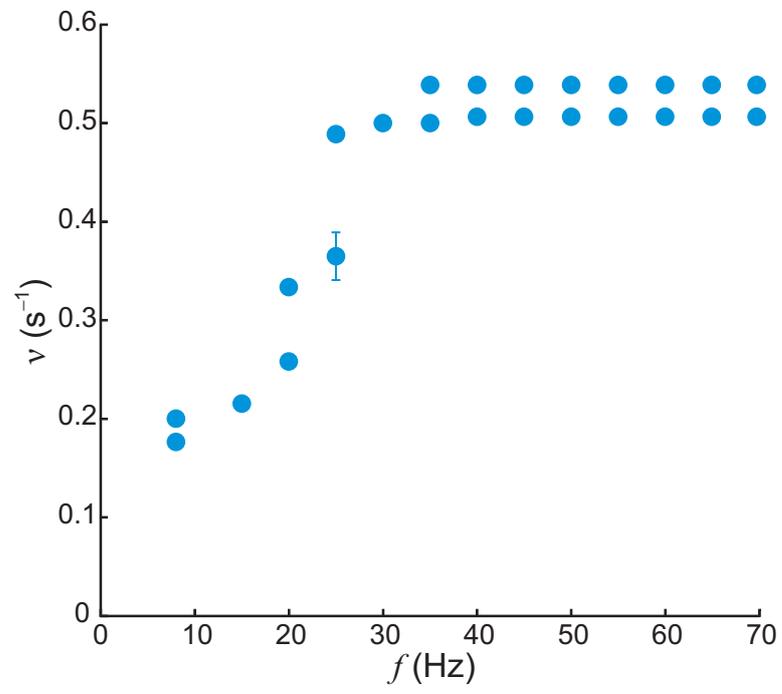

FIG. 8.

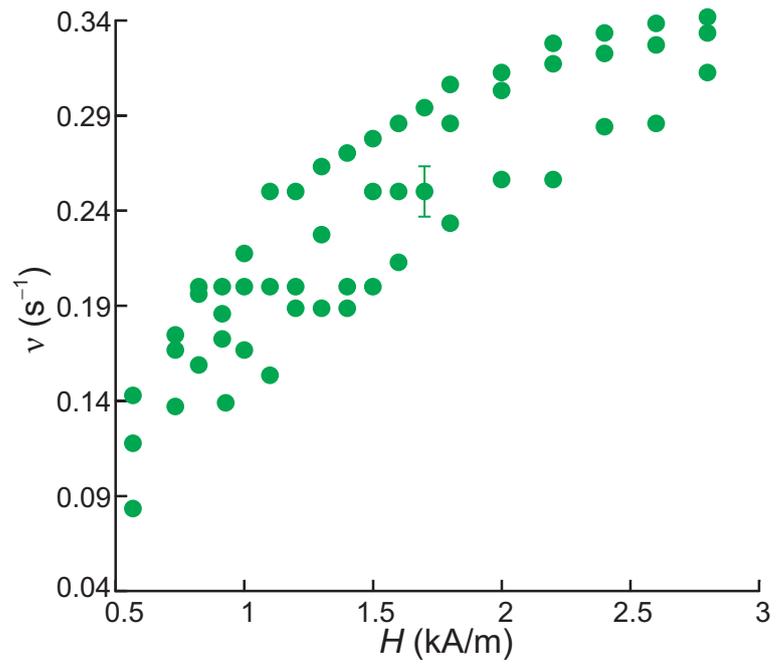

FIG. 9.

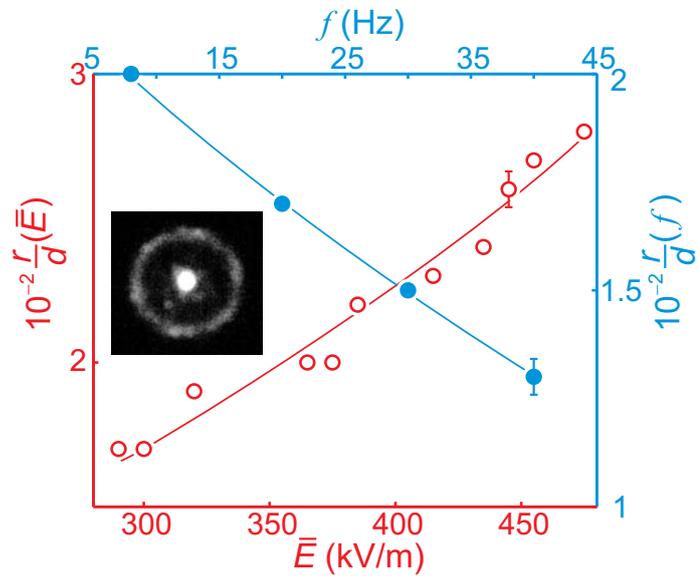

FIG. 10.

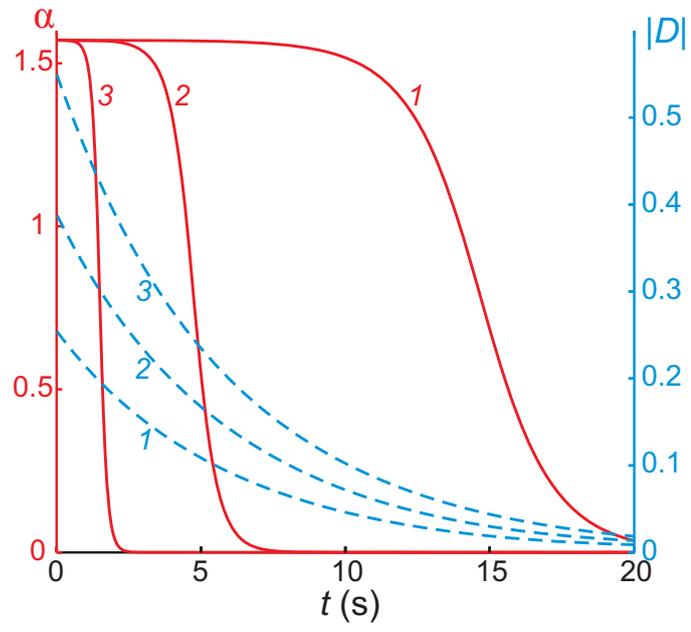

FIG. 11.